\documentclass[12pt]{article}
\usepackage{ifpdf}
\usepackage[a4paper,top=3.2cm,width=17cm, bottom=3.2cm, left=2cm]{geometry}
\usepackage{microtype}
\usepackage[utf8x]{inputenc}
\usepackage[T1]{fontenc}
\usepackage{ae}
\usepackage{aecompl}
\usepackage{setspace}
\usepackage{amsmath}
\usepackage{amsthm}
\usepackage{amssymb}
\usepackage{amsfonts}
\usepackage{dsfont}
\usepackage{mathrsfs}
\usepackage{graphicx} 
\usepackage{cite}
\usepackage{float}

\def\bfseries{\fontseries \bfdefault \selectfont \boldmath}

\usepackage[english]{babel}
\usepackage{datetime}
\shortdate 
\usepackage{bm}
\usepackage{subfigure}
\usepackage{amsmath}

\usepackage[normalem]{ulem} 
\usepackage{soul} 
\usepackage[titletoc,title]{appendix}
\usepackage{authblk}
\usepackage{xcolor,soul}
\usepackage[colorlinks=true,linktocpage=true,linkcolor=blue,citecolor=blue]{hyperref}
\usepackage{cleveref}
\usepackage{todonotes}

\usepackage{setspace}

\newcommand{\pd}{\partial}
\usepackage{titlesec}

\titleformat*{\section}{\large\bfseries}
\titleformat*{\subsection}{\bfseries}
\titleformat*{\subsubsection}{\bfseries}

\newcommand{\be}{\begin{equation}}
\newcommand{\ee}{\end{equation}}
\newcommand{\bea}{\begin{eqnarray}}
\newcommand{\eea}{\end{eqnarray}}
\newcommand{\nn}{\nonumber}

\begin{document}

\title{{\bf Instability of BTZ  black holes in \\ parity-even massive gravity}}

\renewcommand\Authfont{\scshape\small}

\author[1]{Ghadir Jafari}
\author[2,3]{Hesam Soltanpanahi}

\affil[1]{School of Particles and accelerators , Institute for Research in Fundamental Sciences (IPM),
P.O.Box 19395-5531, Teheran, Iran}
\affil[2]{School of Physics and Telecommunication Engineering,
South China Normal University, Guangzhou 510006, China}

\affil[3]{School of Physics, Institute for Research in Fundamental Sciences (IPM),
P.O.Box 19395-5531, Teheran, Iran}

\date{}
\maketitle
\thispagestyle{empty}

\begin{abstract}

We investigate the linearized  equations of motion in three dimensional Born-Infeld gravity theory. 
Motivated by this model, we calculate the quasinormal modes of BTZ black hole solutions of parity-even gravity theories in three dimensions by using numerical methods.
The  results are classified in three families and they are so accurate such that it allows us to propose analytical form for the quasinormal frequencies.
We find new quasinormal modes which have been missing in the literature of the analytical studies of three dimensional massive gravitons.
These new modes do not have  the known tower structure  and they are purely imaginary for any value of the angular momenta.
Considering the complete set of the quasinormal modes, we show that the BTZ black hole solutions are unstable for any value of the parameters of the  theory. We  confirm our numerical results by computing the new eigenmodes analytically at zero angular momentum. 

\end{abstract}
\newpage
\tableofcontents

\section{Introduction}

It is well known that three dimensional Einstein gravity has no propagating degrees of freedom in the bulk. 
In spite of that, Topological Massive Gravity (TMG) \cite{Deser:1981wh}, New Massive Gravity (NMG) \cite{Bergshoeff:2009hq} and
its generalization \cite{Sinha:2010ai,Afshar:2014ffa} and Born-Infeld (BI)  gravity \cite{Gullu:2010pc} have provided a framework to study three dimensional gravity with higher curvature terms which allow massive graviton excitations.
This setup may eventually helps us to  understand the quantum gravity in three dimensions. 
Investigating the quasinormal modes (QNM) of the classical solutions of these theories may teach us about the characteristic features of the corresponding geometries.
For reviews on QNMs of black holes see e.g. \cite{Berti:2009kk, Konoplya:2011qq}.

From  a holographic point of view,  the AdS$_3$/CFT$_2$  correspondence has been well studied in last three decades initiated by Brown and Henneaux \cite{Brown:1986nw}. On the other hand, 
it is known that adding higher derivative curvature terms in the classical bulk gravity theory is dual to going away from strongly coupled regime of the boundary gauge theory.
Therefore, the BI-action, which can be considered as an infinite set of higher derivative terms, is a convenient framework to study the effects of the higher derivatives in the bulk theory,
 and maybe, to investigate the intermediate coupling behaviour, all the way to weakly coupled regime in the dual boundary theory.

In this note we focus on  three dimensional BI-gravity and its QNMs structure. 
We show that the  fourth order  linearized equations of motion { can be factorized into two } second order differential equations, corresponding to {  massless and}  massive graviton excitations,
and has similar form as the QNM equations of the NMG studied in \cite{Myung:2011bn}.
This motivates us to study the QNMs in a more general framework where the results can be extended to other parity-even theories.

The massive mode excitation in  TMG  satisfies a first order differential equation which allows an analytical approach to find the quasinormal frequencies \cite{Sachs:2008gt}.
In \cite{Li:2008dq} it was shown that the second order differential equation of the perturbations in parity-even theories could be decomposed into two first order differential equations. 
Then following the same approach as \cite{Sachs:2008gt} the analytical solutions are computed, which are obviously solutions to the original equation. 
Nevertheless, there is no guarantee that calculated modes are the full solutions of the original second order differential equation.
To cover this missing point in the literature of the QNMs in parity-even 3D gravities, 
we employ a numerical approach to compute the quasinormal modes of the original equation.

Our results show new features of the excitations in three dimensional parity-even massive gravities. 
While our numerical QNMs include the analytical ones, we found new modes which do not appear in the analytical studies.
Due to existence of the new  modes, we show that static BTZ black hole solutions of these theories are unstable for any value of the parameters of the theory and the background.
On top of that, the new modes we found are unique and there is no tower of them in the QNM structure which means they do not belong to any Christmas  tree structure of the modes. In zero angular momentum we could find analytical solution corresponding to the new mode which,  confirms our numerical analysis.

The organization of the paper is as follows. 
In the next section, \ref{3D-gravity} we will shortly review the BI-gravity and its relation with NMG and its extensions to  $\mathcal{O}(R^3)$. 
We show a regime of the parameters which admits the BTZ black hole solutions and we study its boundary theory. 
In section \ref{QNM-eq}, we find the linearized equation of the metric perturbations and investigate the mass of graviton excitations. 
Section \ref{analytic-QNM} is devoted to analytical attempts to find the massive QNMs. 
The heart of our work is section \ref{numeric-QNM} in which we use numerical tools to compute the QNMs. 
In this section we start with vanishing angular momentum and then we turn it on for the more general situation. 
For the former case we support our result by using analytical investigation. 
We close the paper by a summary and outlook in section \ref{discussion}. 
For completeness appendixes \ref{Btensor} and \ref{knotzero}  contain some technical details of the computation of boundary stress tensor and  the coupled QNM equations at finite angular momentum, respectively.

\section{3D-BI gravity}\label{3D-gravity}

In this section we provide a review on BI gravity in three dimension and, its relation with NMG theory \cite{Bergshoeff:2009hq} and its extensions \cite{Sinha:2010ai,Afshar:2014ffa}.
Although the equations of motion of this theory { have a} diversity  of solutions \cite{Ghodsi:2010ev,Gurses:2011fv,Gurses:2012db,Altas:2015dfa}, we will focus on the BTZ geometry and investigate its  physical properties.
The BI extension of NMG introduced in \cite{Gullu:2010pc} as,
\begin{equation}\label{BIG}
I_{\text{BI}}=-\frac{4m^2}{\kappa_3^2}\int d^3x\sqrt{-\det g}\bigg[\sqrt{\det(\mathbf{1}+\tfrac{\sigma}{m^2}g^{-1}G)}-\lambda\bigg]~,
\end{equation}
where $ G_{\mu\nu}=R_{\mu\nu}-\tfrac12 R g_{\mu\nu} $ is the Einstein tensor, $m^2$ is a positive definite dimension-full parameter, $\sigma=\pm1$ is a sign coefficient, $\lambda$ is a parameter related to the cosmological constant and  $\kappa_3$ is related to  the three dimensional Newton constant as $\kappa_3^2=16\pi G_3$. Note that $\lambda=0$ is excluded from the parameter space since at this point the field redefinition $g_{\mu\nu}\to g_{\mu\nu}-\tfrac{\sigma}{m^2}G_{\mu\nu}$ makes the theory trivial \cite{Alkac:2018whk}.
Expanding this action in terms of  $1/m^2$, which is practically { a} derivative expansion, has the following interesting feature
\begin{align}\label{NMG}
I_{\text{NMG}}=-\frac{4m^2}{\kappa^2}\int d^3x\sqrt{-\det g}\big[(1-\lambda) -  \frac{\sigma}{4 \mathit{m}^2} R -  \frac{1}{32 \mathit{m}^4} (8 R_{\alpha \beta} R^{\alpha \beta} - 3 R^2)+O(R^3)\big]~.
\end{align}
The zeroth order expansion gives the Einstein-Hilbert action with a  cosmological constant $\Lambda=\frac{2m^2(1-\lambda)}{\sigma}$, and to first order it reproduces the NMG action \cite{Bergshoeff:2009hq} which is a minimal parity preserving massive gravity in three dimension.
On top of that, it is straightforward to show that expansion of the BI action \eqref{BIG} up to second order reproduces the extension of NMG theory proposed in \cite{Sinha:2010ai}  by demanding the existence of a  holographic c-theorem. This is a fascinating agreement between two independent approaches of investigating higher derivative corrections to Einstein-Hilbert action in three dimensional gravity. 
While direct extensions of NMG to higher order seems to be an extremely difficult task, one may expand the BI action \cref{BIG} to any order of derivative curvature invariants. 
One main difference between BI theory and any truncated version of that appears in their vacuum solutions. 
Although BI has a single vacuum solution, truncated models may have several vacua.\footnote{We  thank B. Tekin for bringing this point to our attention.}

We would like to study the quasinormal modes  of the BTZ black holes of the BI theory by solving the linearized equations of motion. The following formulas  are useful for analytical calculation,
\begin{equation}\label{matrixrules}
\delta(\mathbf{A}^{-1})=-\mathbf{A}^{-1}\cdot\delta\mathbf{A}\cdot\mathbf{A}^{-1},\quad \delta(\sqrt{\det \mathbf{A}})=\tfrac12\sqrt{\det \mathbf{A}}\, \, \, \text{Tr}(\mathbf{A}^{-1}\delta\mathbf{A})~,
\end{equation}
where $ \mathbf{A}$ is a general invertible matrix and  $ \mathbf{A}.\mathbf{A}^{-1}=\mathbf{1} $. 
Looking at  action \eqref{BIG} one may identify a matrix as $ \mathbf{A}^{\mu}{}_{\nu} =\delta^{\mu}{}_{\nu} +\tfrac{\sigma}{m^2}G^{\mu}{}_{\nu}  $, and its variation is given by $ \delta \mathbf{A}^{\mu}{}_{\nu} =\tfrac{\sigma}{m^2}\delta G^{\mu}{}_{\nu}  $.
In order to write the equations of motion it is convenient to define a tensor,
\begin{equation}\label{Vdef}
\mathcal{V}^{\mu}{}_{\nu}:=\sqrt{\det(\mathbf{1}+\tfrac{\sigma}{m^2}g^{-1}G)}(\frac{1}{\mathbf{1}+\tfrac{\sigma}{m^2}g^{-1}G})^{\mu}{}_{\nu}~,
\end{equation}
and  an operator,
\begin{align}\label{Pdef}
P_{\mu\nu\alpha\beta}:=&g_{\mu\nu}R_{\alpha\beta}-g_{\alpha\mu}g_{\beta\nu}(R+\square)\nonumber\\&+g_{\beta\mu}\nabla_{\alpha}\nabla_{\nu }+g_{\beta\nu}\nabla_{\alpha}\nabla_{\mu}-g_{\mu\nu}\nabla_{\alpha}\nabla_{\beta}
+g_{\mu\nu}g_{\alpha\beta}\square-g_{\alpha\beta}\nabla_{\mu}\nabla_{\nu}~.
\end{align}
Using the following relation for variation of Einstein tensor,
\begin{align}\label{deltaG}
\delta G_{\mu\nu}=\tfrac12 P_{\mu\nu\alpha\beta}\delta g^{\alpha\beta}~,
\end{align}
one may show that the  equations of motion have the following compact form,
\begin{equation}\label{EOM}
\frac{\sigma}{2m^2}P_{\mu\nu\alpha\beta}\mathcal{V}^{\alpha\beta}  +g_{\mu\nu}\bigg[\sqrt{\det(\mathbf{1}+\tfrac{\sigma}{m^2}g^{-1}G)}-\lambda\bigg]=0~.
\end{equation}

We are interested in BTZ geometries which are examples of  locally maximally symmetric spacetimes defined by 
\be\label{maximallySym}
R_{\alpha\beta\mu\nu}=\pm\frac{1}{\ell^2}(g_{\alpha\mu}g_{\beta\nu}-g_{\beta\mu}g_{\alpha\nu})
\ee
where $\pm$ corresponds to the dS/AdS respectively, $\ell$ is the radius of dS/AdS and for flat spacetime the right hand side vanishes. 
From now on, we focus on Anti-de Sitter solutions and we work with minus sign in \cref{maximallySym}. For AdS geometries
 one may show that the tensor defined in \cref{Vdef} simplifies to
\begin{equation}\label{Vads}
\mathcal{V}^{\mu\nu}=\sqrt{1 + \tfrac{\sigma }{\ell^2 \mathit{m}^2}}\  g^{\mu\nu}
\end{equation}
Plugging the above results  in the equation of motion \eqref{EOM} one may obtain a relation between the AdS radius $\ell$ and the parameters of the theory as
\begin{equation}\label{lamda3}
\lambda=\sqrt{1+\tfrac{\sigma}{m^2\ell^2}}
\end{equation}
This shows that for given parameters of the theory, there is a unique vacuum solution.
Expanding this equation to second order in $ 1/m^2$, one may find,  
\[ \lambda=1+\tfrac{\sigma}{2\ell^2 \mathit{m}^2}-\tfrac1{8\ell^4m^4}+\mathcal{O}(1/m^6)  \]
which is the corresponding relation for vacua geometries in NMG theory \cite{Bergshoeff:2009hq}.
Note that in order to have locally AdS solution, satisfying \cref{lamda3} , there are two choices for parameters of the theory with $m^2>0$, namely
\begin{equation}\label{solcondition}
\begin{cases}
\sigma=-1  \ \ \ \& & 0<\lambda<1\\
\ \ \ \mathrm{Or}   \\
\sigma=1 \ \  \ \ \ \& &  \lambda>1
\end{cases}.
\end{equation}
So far we have reviewed some features of the 3D-BI theory under conditions of having AdS solution.
According to the holography conjecture, one may expect that there is a boundary theory dual to an asymptotically AdS background. 
The gravity side of the theory is a re-summation  of infinite number of higher derivative corrections to the Einstein-Hilbert action.
Therefore the field theory side might be related to weakly coupled regime of a dual gauge theory. 
In the following subsection we investigate this boundary theory by computing the corresponding central charge and the stress tensor.

 \subsection{Boundary theory data}

Boundary theory of AdS$_3$ geometries has been investigated in last 3 decades initiated by Brown and   Henneaux \cite{Brown:1986nw}. 
In this section we  make some comments from this perspective on the boundary theory of BTZ black holes in three dimensional BI-gravity.
In three dimensions the central charge for higher derivative gravities can be obtained by using the Wald formula \cite{Wald:1993nt,Saida:1999ec}
 \begin{equation}\label{ccharge}
c=\frac{\ell}{2G_3}g_{\mu\nu}\frac{\pd\mathcal{L}}{\pd R_{\mu\nu}}~,
\end{equation}
where $G_3$ is three dimensional Newton constant and $\mathcal{L}$ is the Lagrangian density.
Since the BI action \eqref{BIG} is introduced as a function of Einstein tensor we use the following chain rule,
\begin{equation}
\frac{\pd\mathcal{L}}{\pd R_{\mu\nu}}=\frac{\pd\mathcal{L}}{\pd G_{\alpha\beta}} \frac{\pd G_{\alpha\beta}}{\pd R_{\mu\nu}}=\frac{\pd\mathcal{L}}{\pd G_{\alpha\beta}}(\delta^\mu_\alpha\delta^\nu_\beta-\tfrac12g^{\mu\nu}g_{\alpha\beta})
\end{equation}
to express the derivative in \cref{ccharge} in terms of derivative with respect to Einstein tensor. 
By employing  \cref{matrixrules} and the tensor defined in \cref{Vdef} one may show that
\begin{equation}
\frac{\pd\mathcal{L}}{\pd G_{\alpha\beta}}=2\sigma \mathcal{V}^{\alpha\beta}=2\sigma\lambda g^{\alpha\beta},
\end{equation}
where in the second equality we plug the conditions for maximally symmetric solution \cref{Vads,lamda3}.
Using these formulas the corresponding central charge  simplifies to
\begin{equation}\label{ccharge2}
c=\frac{3\ell}{2G_3}\,\sigma\lambda~.
\end{equation}
This is nothing but the scaling of Brown-Henneaux central charge \cite{Brown:1986nw} computed for Einstein theory.
This expression has been found in \cite{Gullu:2010st,Alishahiha:2010iq} and it matches the Weyl anomaly computation, but here we could rederive it with less difficulty. 

From holographic point of view one may expect that the central charge \eqref{ccharge2} corresponds to the zero energy of a dual (conformal) field theory.
Unitarity of the theory requires that this central charge should be non-negative which leads to,
\be\label{sigma-lambda}
\sigma\lambda\geq0~.
\ee
On the other hand, we found a regime of the parameters of the BI action \eqref{BIG}  in which the BTZ geometries are solutions to the equations of motion  in \cref{solcondition}. 
Now by comparing two conditions in  \cref{solcondition,sigma-lambda}, 
one can see that the unitarity of the boundary theory fixes the sign parameter  $\sigma$ and restricts the regime of $\lambda$, namely\footnote{Note that $\lambda=0$ leads to a trivial theory as mentioned previously.}
\begin{equation}\label{solcondition1}
\sigma=1,\qquad  \lambda>1~.
\end{equation} 

In order to find the correct energy of a solution in a theory one may follow the Hamilton-Jacobi analysis\cite{Brown:1992br}. 
The first step in this approach is  calculating the correct boundary terms so that the variational principle is well-posed. 
For general metric backgrounds boundary terms are known for a few number of models  such as Lovelock theories \cite{PhysRevD.36.392}. 
But for a symmetric background the boundary terms can be computed in a systematic way. 
In  \cref{Btensor} we show that for a maximally symmetric geometry in BI theory the boundary term is given by
\begin{equation}\label{GHYtems}
\mathcal{L}_{\text{boundary}}=2\frac{\sigma\lambda}{\kappa_3^2}\int d^2x \sqrt{|\hat{h}|} K~,
\end{equation}
which leads to the Brown-York stress tensor as
\be\label{stress-tensor}
T^{ab}=\sigma\lambda (K^{ab}-\hat{h}^{ab}K)~,
\ee
in which $\hat{h}_{ab}$ is the boundary metric and $K_{ab}$ is the extrinsic curvature.
Interestingly enough, comparing this stress tensor with the results for Einstein theory one see the same scaling factor $(\sigma\lambda)$ as  found for the central charge in \cref{ccharge2}.
Therefore, the energy density computed from the time-time component of the energy-momentum tensor \eqref{stress-tensor} 
does not modify the regime of the parameters.

Similar to Einstein theory we can compute the central charge from the energy-momentum tensor given \cref{stress-tensor}, which obviously leads to \cref{ccharge2},  and this  could be considered as a crosscheck for our analysis.
As a side remark we would like to note that, similar to our discussion after \cref{lamda3}, 
the central charge and Brown-York stress tensor for NMG theory and its $n$th order extended versions can be found  by expanding our results for BI action { to} $\mathcal{O}(1/m^2)$ and $\mathcal{O}(1/m^{2n})$ respectively.

\section{Linearized equations of motion}\label{QNM-eq}

In this section we  linearize  equations of motion \eqref{EOM}, derived in previous section for BI gravity, and compare its expansion in $1/m^2$ 
with the linearization of NMG and extended NMG. 
 We consider the metric  as $ g_{\mu\nu}=\bar{g}_{\mu\nu}+h_{\mu\nu} $, in which  $ \bar{g}_{\mu\nu} $ is the background spacetime and $h_{\mu\nu}$ is the perturbation on top of that.
Using the rules  in  \cref{matrixrules} and giving some massage, the equations of motion for perturbation is found
to be,
\begin{align}\label{BI-pert}
&\frac{1}{\sqrt{1+\tfrac{\sigma}{\ell^2m^2}}}\big[\tfrac{2}{\ell^2} (1+  \tfrac{2\sigma}{\ell^2m^2} ) \mathit{g}_{\mu \nu} \tilde{\mathit{h}} - \tfrac{4}{\ell^2} (1 +  \tfrac{3\sigma}{\ell^2m^2}) \tilde{\mathit{h}}_{\mu \nu} - 2(1+\tfrac{5\sigma}{\ell^2m^2}) \square\tilde{\mathit{h}}_{\mu \nu} \nonumber\\ &-2 (1-\tfrac{2\sigma}{\ell^2m^2} ) \mathit{g}_{\mu \nu} \nabla_{\beta}\nabla_{\alpha}\tilde{\mathit{h}}^{\alpha \beta}+ (1 +  \tfrac{2\sigma}{\ell^2m^2}) \mathit{g}_{\mu \nu} \square\tilde{\mathit{h}} +4 (1 - \tfrac{\sigma}{\ell^2m^2}  ) \nabla_{(\mu}\nabla_{\alpha}\tilde{\mathit{h}}_{\nu)}{}^{\alpha} -  \nabla_{\nu}\nabla_{\mu}\tilde{\mathit{h}}  \nonumber\\&+\tfrac{\sigma}{m^2} \big(4  \nabla_{(\mu}\square\nabla_{\alpha}\tilde{\mathit{h}}_{\nu)}{}^{\alpha} - 2  \nabla_{(\mu}\square\nabla_{\nu)}\tilde{\mathit{h}} -   \mathit{g}_{\mu \nu} \square\nabla_{\beta}\nabla_{\alpha}\tilde{\mathit{h}}^{\alpha \beta} \nonumber\\& -   \nabla_{\nu}\nabla_{\mu}\nabla_{\beta}\nabla_{\alpha}\tilde{\mathit{h}}^{\alpha \beta} +  \nabla_{\nu}\nabla_{\mu}\square\tilde{\mathit{h}}- 2  \square^2\tilde{\mathit{h}}_{\mu \nu} +  \mathit{g}_{\mu \nu} \square^2\tilde{\mathit{h}}\big)\big]=0~,
\end{align}
{with}  $ \tilde{h}_{\mu\nu}:=h_{\mu\nu}-g_{\mu\nu}h $. 
The pre-factor in this equation plays a crucial role later when we want to compare with the results for NMG theories by expanding \cref{BI-pert} in $1/m^2$ and therefore, we keep it to the end. 
The trace of above equation has a simple expression,
\begin{equation}\label{BI-pert-trace}
\frac{1}{\sqrt{1+\tfrac{\sigma}{\ell^2m^2}}} (\nabla_{\mu }\nabla_{\nu}\tilde{h}^{\mu\nu}-\tfrac{1}{\ell^2} \tilde{h})=0~.
\end{equation}
On the other hand, the Harmonic gauge $ \nabla_{\mu }\tilde{h}^{\mu\nu}=0 $ with the aid of linearized equation of motion \eqref{BI-pert-trace}, leads to  the so called transverse traceless (TT) gauge for original metric perturbations, $\nabla_{\mu }h^{\mu\nu}=h=0 $.
In this gauge,  the linearized equation of motion simplifies to,
\begin{equation}\label{key}
\frac{1}{\sqrt{1+\tfrac{\sigma}{\ell^2m^2}}} \big[-\tfrac{2}{\ell^2} (1 + 3\tfrac{\sigma}{\ell^2m^2} ) \mathit{h}_{\mu \nu} - ( 1+\tfrac{5\sigma}{\ell^2m^2}) \square\mathit{h}_{\mu \nu} - \tfrac{\sigma}{m^2} \square^2\mathit{h}_{\mu \nu}\big]=0~,
\end{equation}
which, equivalently,   can be factorized as,
\begin{equation}\label{pereq1}
\frac{1}{\sqrt{1+\tfrac{\sigma}{\ell^2m^2}}}(\square+\sigma m^2+\tfrac{3}{\ell^2})(\square+\tfrac{2}{\ell^2}) h_{\mu\nu}=0.
\end{equation} 
Therefore the equation of motion for linear perturbation splits into two parts. 
Similar structure has been found for TMG \cite{Sachs:2008gt} and NMG \cite{Myung:2011bn} theories.
The second parenthesis in \cref{pereq1}  is related to a massless graviton in BTZ geometry while, the first one corresponds to a massive graviton with the mass
\begin{equation}\label{gravitonmassBI}
\mathbf{m}^2=-\sigma m^2(1+\tfrac{\sigma}{m^2\ell^2})=-\sigma m^2\lambda^2.
\end{equation}
Note that for $ \sigma=-1 $, which corresponds to non-unitary boundary theory \eqref{solcondition1}, there is a critical point at $ m \ell=1 $  where the mass of graviton, the parameter $\lambda$ and the central charge of the boundary theory \eqref{ccharge2} vanish but, at this critical point the BI-model becomes trivial.
However,  for $ \sigma=-1 $ and $m \ell\neq1$ the mass of gravioton satisfies the Breitenlohner Freedman bound $\mathbf{m}^2>m_{\text{BF}}^2=-1/\ell^2$ \cite{Breitenlohner:1982bm}.
On the other hand,  $\sigma=1$, which gurantees the unitarity of the boundary theory, leads to  BF-bound  violation, $\mathbf{m}^2<m_{\text{BF}}^2$.
This indicates the "bulk-boundary clash" \cite{Bergshoeff:2009aq} of  BI-model mentioned in \cite{Gullu:2010vw} (for a review see  \cite{Merbis:2014vja}). 

At this point, we would like to discuss about how one can get the NMG data from our results. 
Expanding \cref{pereq1}, including the pre-factor,  to the first order in $ 1/{m^2} $ leads to 
\begin{equation}\label{perNMG}
  (\square+\sigma m^2+\tfrac{5}{2\ell^2}+\mathcal{O}(\tfrac{1}{m^{2}}))(\square+\tfrac{2}{\ell^2})h_{\mu\nu}=0~,
\end{equation}
which  is the equation for perturbations in NMG theory\cite{Myung:2011bn} and apparently  the pre-factor has an important role in the first parenthesis.
One may calculate the graviton mass for NMG theory by looking at the first term in \cref{perNMG} which is given by
\begin{equation}\label{gravitonmassNMG}
\mathbf{m}_{\text{NMG}}^2=-\sigma m^2(1+\tfrac{\sigma}{2m^2\ell^2}).
\end{equation}
Now the expansion of  \eqref{pereq1} to  second order in $ 1/{m^2} $  leads to
\begin{align}\label{perENMG}
\left[ \left(1-\tfrac{\sigma }{2 l^2 m^2}\right)\square+\sigma m^2+\tfrac{5}{2 l^2}-\tfrac{9 \sigma }{8 l^4 m^2}+\mathcal{O}\left(\tfrac{1}{m^{4}}\right)\right](\square+\tfrac{2}{\ell^2}) h_{\mu\nu}&=\nn\\
 \left(1-\tfrac{\sigma }{2 l^2 m^2}\right)\left[\square+\sigma m^2+\tfrac{3}{\ell^2}+\tfrac{3 \sigma}{8m^2\ell^4}+\mathcal{O}(\tfrac{1}{m^{4}})\right](\square+\tfrac{2}{\ell^2}) h_{\mu\nu}&=0~.
\end{align}
Note that in the second line after factoring out the coefficient of box operator, we expand the remaining terms in $1/m^2$ up to the second order.
Using  this linearized  equation  the mass of graviton is given by
\begin{equation}\label{gravitonmassENMG}
\mathbf{m}_{\text{NMG}^{(2)}}^2=-\sigma m^2(1+\tfrac{\sigma}{m^2\ell^2}+\tfrac{3}{8m^4\ell^4}).
\end{equation}
By following this procedure of expansion  to $\mathcal{O}(1/m^{2n})$ one can show that  expression for the mass of graviton has an interesting structure
\begin{equation}\label{gravitonmassOn}
\mathbf{m}_{\text{NMG}^{(n)}}^2=-\sigma m^2\left(1+\tfrac{\sigma}{m^2\ell^2}+\tfrac{c_n}{(\ell^2m^2)^n}\right).
\end{equation}
where $ c_n$'s are numerical coefficients and their absolute value are less than one, $|c_n|<1$.  
Note that, apart from the first two terms in \cref{gravitonmassOn},  there is only one other term, which is proportional to  ${1}/({m^2\ell^2})^n$ and
at the limit $ n\to\infty $, this  term would disappear if we demand $m\,\ell>1$.
Let us emphasize that, all the potentially intermediate terms in the expansion \eqref{gravitonmassOn} disappear due to the similar factoring procedure used in \cref{perENMG}.
This is a fascinating result which shows how the truncation of the expansion in $1/{m^2} $ at any order, which leads to three terms in right hand side of \cref{gravitonmassOn}, would be different than keeping all the terms as in BI theory, which corresponds to \cref{gravitonmassBI} and  is in agreement with the first two terms in \cref{gravitonmassOn}.

\section{Analytical analysis of QNM's}\label{analytic-QNM}

In this section we are following \cite{Sachs:2008gt}  to investigate the quasinormal frequencies of the BTZ black hole in BI theory.
The massive mode equation given in  \cref{pereq1} fits in a more general equation namely,
\begin{equation}\label{2ordereq}
(\square+a)h_{\mu\nu}=0~,
\end{equation}
where parameter $a$ depends on the details of the theory and the background data.\footnote{In our analysis we don't consider the $a=2/\ell^2$ which corresponds to the critical point discussed after \cref{gravitonmassBI}.} 
In this way, we can compare the massive QNM's of BTZ black holes in different theories. Specially we want to investigate the effect of truncation of BI theory at a given order of expansion in $1/{m^2}$.

In fact in three dimension one may further decompose \cref{2ordereq} to two first order equations. 
To  this aim, consider the following operator
\begin{equation}\label{Dd}
\mathcal{D}_{\mu\nu}^{(M)}=\varepsilon_{\mu\nu\rho}\nabla^\rho+ \tfrac{M}{\ell}\ g_{\mu\nu}
\end{equation}
with a constant $ M $. Using this operator it is easy to show that
\begin{equation}\label{D2}
\left(\mathcal{D}^{(-M)}\mathcal{D}^{(M)}\right)_\mu^{~\rho} h_{\rho\nu}=\square\mathit{h}_{\mu \nu}+(\tfrac{3-M^2}{\ell^2}) \mathit{h}_{\mu \nu}  -  \nabla_{\mu}\nabla_{\beta}\mathit{h}_{~\nu}^{\beta}-\tfrac{\mathit{h}}{\ell^2} \mathit{g}_{\mu \nu}~.
\end{equation}
 In TT-gauge this relation is equivalent to  \cref{2ordereq} by setting $ M=\sqrt{3-a\ \ell^2} $ . Therefore every solution of the equation
\begin{equation}\label{1ordereq}
\epsilon_\alpha{}^{\mu\nu}\nabla_\nu h_{\mu\beta}\pm\tfrac{M}{\ell}\ h_{\alpha\beta}=0
\end{equation} 
is a solution of the second order  \cref{2ordereq}.
It is easy to show that the TT-gauge condition could be deduced from \cref{1ordereq}. 
But one must note that there is a critical point where $M=0$ at which, logarithmic solutions may appear. 
We do not consider this case in our analysis while it has been studied in various cases \cite{Sachs:2008yi,Grumiller:2008qz,Grumiller:2009sn}.

In order to solve  \cref{1ordereq} one may consider the global coordinate system for BTZ black hole,
\begin{equation}\label{GCBTZ}
ds^2=\ell^2\left(-\sinh^2\rho dt^2+\cosh^2\rho d\phi^2+d\rho^2\right)~,
\end{equation}
such that the left/right moving solutions are given by \cite{Sachs:2008gt}
\begin{eqnarray}\label{right}
h^{r}_{\mu\nu}=e^{-ik(t+\phi)} e^{-2h_r
	t}(\sinh\rho)^{-2h_r}(\tan\rho)^{-ik}\left(
\begin{array}{ccc}
1 & 0 & \frac{2}{\sinh2\rho} \cr 0 & 0 & 0 \cr \frac{2}{\sinh2\rho}
& 0 & \frac{4}{\sinh^22\rho}
\end{array}
\right)\,,~~~h_r=\frac{1}{2}(\mp M-1)~, \label{sol1}
\end{eqnarray}
and
\begin{eqnarray}\label{left}
h^{l}_{\mu\nu}=e^{-ik(t+\phi)}
e^{-2h_lt}(\sinh\rho)^{-2h_l}(\tan\rho)^{ik}\left(
\begin{array}{ccc}
0 & 0 & 0 \cr 0 & 1 & \frac{2}{\sinh2\rho} \cr 0 &
\frac{2}{\sinh2\rho} & \frac{4}{\sinh^22\rho}
\end{array}
\right)\,,~~~h_l=\frac{1}{2}(\pm M-1)~. \label{sol2}
\end{eqnarray}
The ingoing boundary condition at the horizon selects the positive signs in $ h_{r,l}$ and therefore  the least damped quasinormal frequencies are ,
\begin{eqnarray}
&&\omega^r_{n}=-2ih_r+k=-i(M-1)+k~,\\
&&\omega^l_{n}=-2ih_l-k=-i(M-1)-k~,
\end{eqnarray}
where  periodicity along $\phi$ leads to discrete angular momenta $k$.

Introducing the operator $ (L_{-1}\bar{L}_{-1}) $,  where $ L_{k} $ are Killing vectors associated with background BTZ solution, one may find an infinite tower of quasinormal modes,
\begin{equation}\label{nQM}
h^{(n)}_{\mu\nu}=(L_{-1}\bar{L}_{-1})^n h_{\mu\nu}~,
\end{equation}
and the corresponding frequencies with overtone number $n$,
\be\label{QNMs}
\omega_{n}=-i(M-1+2n)\pm k
\ee
It is easy to see that  $ M>1 $ leads to stable frequencies. 
There is an important point regarding the solutions \eqref{right} and \eqref{left}. 
As mentioned in \cite{Myung:2011bn}, these two modes are not orthogonal, so they can not be interpreted as independent massive modes in the theory and they might be some missing solutions to the linearized \cref{2ordereq}. We will come back to this point in next section.
 
By comparing  \cref{2ordereq} with the massive mode equation in different theories one may find the relation between the auxiliary positive parameter $M$ and the parameters of the theory.
For  NMG, $ M^2=-\sigma\, m^2+\frac{1}{2\ell^2}$ while for the generalized NMG it will be given by $ M^2=-\sigma\, m^2-\frac{3}{8\ell^4m^2} $. 
The general structure for the truncated BI-action (generalized NMG to higher orders) can be found as
\be\label{ghadir}
M^2_{\text{NMG}^{(n)}}=-\sigma\, m^2\left(1+\tfrac{c_n}{(\ell^2m^2)^n}\right)
\ee
 On the other hand for the BI-gravity, by comparing \cref{pereq1,2ordereq}, it is easy to see
\be
M^2_{\text{BI}}=-\sigma\, m^2,
\ee
which, could be found by taking the limit $n\to\infty$ from the results for extended NMG theory \eqref{ghadir} by demanding $m\,\ell>1$.

At the end of this section we would like to point out the main linearized equation that one has to solve for the BI theory and other parity even massive gravity models is a second order differential equation given in \eqref{2ordereq} while the analytical investigation is based on factorized first order equations \eqref{1ordereq}.
Although it is clear that the solutions of to the first order equations \eqref{1ordereq} are solutions to the second order one \cref{2ordereq}, it is not obvious whether it works other way around. 
In other words, the set of solutions to the later could be larger than the former. 
Of course, the same argument applied when we
factorized the fourth order differential equations to one massless and one massive excitations in \cref{pereq1}.
But, in this work we just focus on the massive mode and not the full linearized equation.
In next section we use numerical tools to tackle linearized equation \eqref{2ordereq} with given boundary conditions corresponding to the quasinormal modes and we will compare the results of these two approaches.

\section{Numerical analysis of QNM's}\label{numeric-QNM}

In this section, by using a numerical method, we compute the quasinomal frequencies of  BTZ black holes as solutions of BI theory or any 3D theory of gravity  which has a massive mode corresponding to  \cref{2ordereq}. 
To this aim, we consider the static BTZ black hole in Eddington-Finkelstein coordinate
\begin{equation}
ds^2=\frac{\ell^2}{r^2}\left[-(1-\frac{r^2}{r_0^2})dt^2-2dtdr+d\phi^2\right]
\end{equation}
where $ \ell $ is the radius of $AdS_3$  spacetime, and the asymptotic region   is $ r\to0 $.
Linearizing equation of motion on BTZ geometry in this coordinate has a twofold advantages. 
First, the ingoing boundary condition at the event horizon is automatically satisfied. Second, the vanishing source boundary condition at the asymptotic boundary (which is the definition of QNM) is easy to impose as we will discuss.

It is common to use the plane wave Ansatz for the fluctuations  and work in momentum space, $h_{\mu\nu}(r,t,\phi)=\mathcal{Z}_{\mu\nu}(r)e^{-i\frac{\omega}{r_0} t+i \frac{k}{r_0}\phi}$. 
The transverse traceless gauge, not only reduces the linearized equations to  \cref{pereq1} but also imposes four constraints on the components of $ h_{\mu\nu} $ which are useful to simplify the QNM equations. 

\subsection{$k=0$}

We start with  zero angular momentum $k=0$ which leads to simpler equations. 
It is easy to show that in Eddington-Finkelstein coordinate there are two sets of metric perturbations which are coupled among each other and decoupled from the other ones. 
These two sets are $(h_{t\phi}, h_{r\phi})$ and  $(h_{rr}, h_{tr}, h_{tt}, h_{\phi\phi})$ which are orthogonal to each other and contain all of the metric perturbations.
But using the constraints of the TT-gauge, one can show that the linearized equation \eqref{2ordereq} will reduce to two decoupled equations for metric perturbations $h_{r\phi}$ and $h_{\phi}$.
At the practical level, the following redefinitions are natural to have proper boundary behaviour,
\begin{align}\label{redefine1}
h_{r\phi}(r,t,\phi)=r^{-1-\sqrt{3-a \ell^2}}\mathcal{Z}_1(r)e^{-i\frac{\omega}{r_0} t+i \frac{k}{r_0}\phi}~,\qquad\qquad h_{\phi\phi}(r,t,\phi)=r^{-1-\sqrt{3-a \ell^2}}\mathcal{Z}_2(r)e^{-i\frac{\omega}{r_0} t+i \frac{k}{r_0}\phi}~.
\end{align} 
It is easy to show that using the following redefinitions
\be\label{redefine2}
M=\sqrt{3-a\,\ell^2}~,\qquad \qquad \frac{r}{r_0}= u~,
\ee
reduce the equations in terms of new functions as
{\small
\begin{align}
&\left[(M-3) (M+1) (M-1)^2 u-i M \omega  \left(-2 M^2+3 (M-3) u^2+3 M-1\right)+3 (2 M-1) u \omega ^2\right]\mathcal{Z}_1(u) \nonumber\\
&+\left[2 i u \omega  \left(-M^2+3 (M-2) u^2-3 M+1\right)+\left(1-M^2\right) \left((2 M-5) u^2-2 M+1\right)-6 u^2 \omega ^2\right] \mathcal{Z}_1'(u)\nonumber\\
&+\left[\left(M^2-1\right) u \left(u^2-1\right)-3 i u^2 \left(u^2-1\right) \omega \right] \mathcal{Z}_1''(u)=0~,\label{Z1-eq}\\
&\bigg[\left(M^2-1\right)^2 u \left(u^2-(M-2) M\right)+\omega ^2 \left(-M^2 \left(u^2+2\right) u+2 M u+u^3\right)\nonumber\\
&+i \left(M^2-1\right) \omega  \left(-2 M^3+M^2+2 M u^2+u^2\right)-i (2 M+1) u^2 \omega ^3\bigg]\mathcal{Z}_2(u)+\bigg[2 i \left(M^2-1\right) u \omega  \left(M^2-u^2\right)\nonumber\\
&+\left(M^2-1\right) \left(-2 M^3+M^2+\left(2 M^3-3 M^2+2 M+1\right) u^2+(1-2 M) u^4\right)+2 i u^3 \omega ^3\nonumber\\
&+u^2 \omega ^2 \left((2 M-1) u^2-2 M-1\right)\bigg] \mathcal{Z}_2'(u)+\left[\left(M^2-1\right) u \left(u^2-1\right) \left(u^2-M^2\right)+\left(u^3-u^5\right) \omega ^2\right] \mathcal{Z}_2''(u)=0~.\label{Z2-eq}
\end{align}
}  
Note that the same parameter $M$ defined in \cref{redefine2} has been used in the analytical studies introduced in \cref{Dd}.
Interestingly enough, not only $M$ is the only  parameter in both equations but also, there is no temperature dependency since we define the frequency and angular momentum in $r_0$ unit. 
Also, it is easy to see that in both equations the near boundary solution to the  \cref{Z1-eq,Z2-eq} is
\begin{equation}\label{near-boundary}
\mathcal{Z}_i(u)=\mathcal{A}_i(1+\mathcal{O}(u))+\mathcal{B}_iu^{2M}(1+\mathcal{O}(u))~.
\end{equation}
From this expansion it is obvious that we are interested in the cases in which  parameter $M$ is positive  which means  $a \ell^2<3$.
In principle, for 3D massive gravity theories the parameter $M$ could be related to the mass of the graviton in BTZ background and the parameters of the theory. 
Note that $M=0$ is not in the domain of our interest since it corresponds to the logarithmic cases  and $M=1$ is associated with  the massless graviton mode. 

According to the definition of quasinormal modes in asymptotically AdS backgrounds, the mode should be source free ($\mathcal{A}_i=0$) and should satisfy the ingoing boundary condition at the horizon.
On the other hand, according to the holography dictionary, the QNMs correspond to the poles of the retarded Green's function of the dual operators \cite{Kovtun:2005ev}. 
Therefore, our studies may shed some light on a field theory dual to the BI-gravity which might be considered as an infinite summation of the higher derivative corrections to the 3D Einstein-Hilbert action.

For our numerical calculation, we use Chebyshev spectral method to  discretize the radial coordinate between the horizon ($u=1$) and the asymptotic boundary ($u=0$) and solve the eigenvalue \cref{Z1-eq,Z2-eq} to find the QNM's. 
This method has been widely used in QNM literature and implementing the method practically has been  reviewed in \cite{Jansen:2017oag}.
We should emphasize that by using this numerical method we are able to find the lowest QNMs such that, the higher the number of grid points the more number of lowest QNMs one could compute with higher accuracy. 
Nonetheless, accurate results may help us to propose analytical structure for the frequencies.

Here we summarize  our numerical results with high precision (in our numerical calculation we use the number of significant decimal digits between 100 to 1000) and large number of grid points along the radial coordinate (between 100 to 250). 
The results are so accurate which shows following analytic forms:

\begin{enumerate}
\item The frequencies for the $\mathcal{Z}_1$ are classified in two sets,
\begin{equation}\label{QNM-analytic-k1}
\omega^{(1,1)}_n=-i\,\left(M-1+2\,n\right)~,\qquad \omega^{(1,2)}_n=-i\,\left(M+3+2\,n\right)~, \qquad n=0, 1, 2, \cdots
\end{equation}
\item The frequencies for the $\mathcal{Z}_2$ are classified in three sets,
\begin{align}\label{QNM-analytic-k}
&\omega^{(2,1)}_n=-i\,\left(M-1+2\,n\right)~,\qquad \omega^{(2,2)}_n=-i\,\left(M+3+2\,n\right)~, \qquad n=0, 1, 2, \cdots~,\nonumber\\
&\omega^{(2,3)}=-i\,\left(1-M^2\right)
\end{align}
\end{enumerate}
Obviously, the first two sets in $\mathcal{Z}_2$ channel  coincide with the two sets of  $\mathcal{Z}_1$ channel and, 
the first set is in perfect agreement with the analytical expressions given in \cref{QNMs} for zero angular momenta.
But here we find  the third set in $\mathcal{Z}_2$ channel with a single mode which has been missing in the analytical investigations.
On top of that, the  degeneracy between the first and second sets for some modes was not apparent in our analytical results.
Note that the dependency of the modes on the parameter $M$ is completely different in the third set which motivate us to put this unique mode in a separated set.
With these results some comments are in order:
\begin{itemize}
\item Since the largest power of $\omega$ in equation \cref{Z1-eq}/\eqref{Z2-eq} is two/three, one may expect that there are two/three sets of QNM's in $\mathcal{Z}_{1/2}$ channel a priory.
\item Except the first two modes in $\omega^{(i,1)}$ set there is a degeneracy in other modes between first and second sets namely, $\omega^{(i,1)}_{n+2}=\omega^{(i,2)}_{n}~~ \text{for}
\quad  n=0, 1, 2, \cdots$.
\item For $0<M<1$, the zeroth mode in the  first family, $\omega^{(i,1)}_0$, is unstable and all other modes are stable.
\item For $M>1$, the single mode in the third family $\omega^{(2,3)}$ is unstable and  all other modes  are stable.
\end{itemize}

\begin{figure}
	\begin{center}
		\includegraphics[height=5cm]{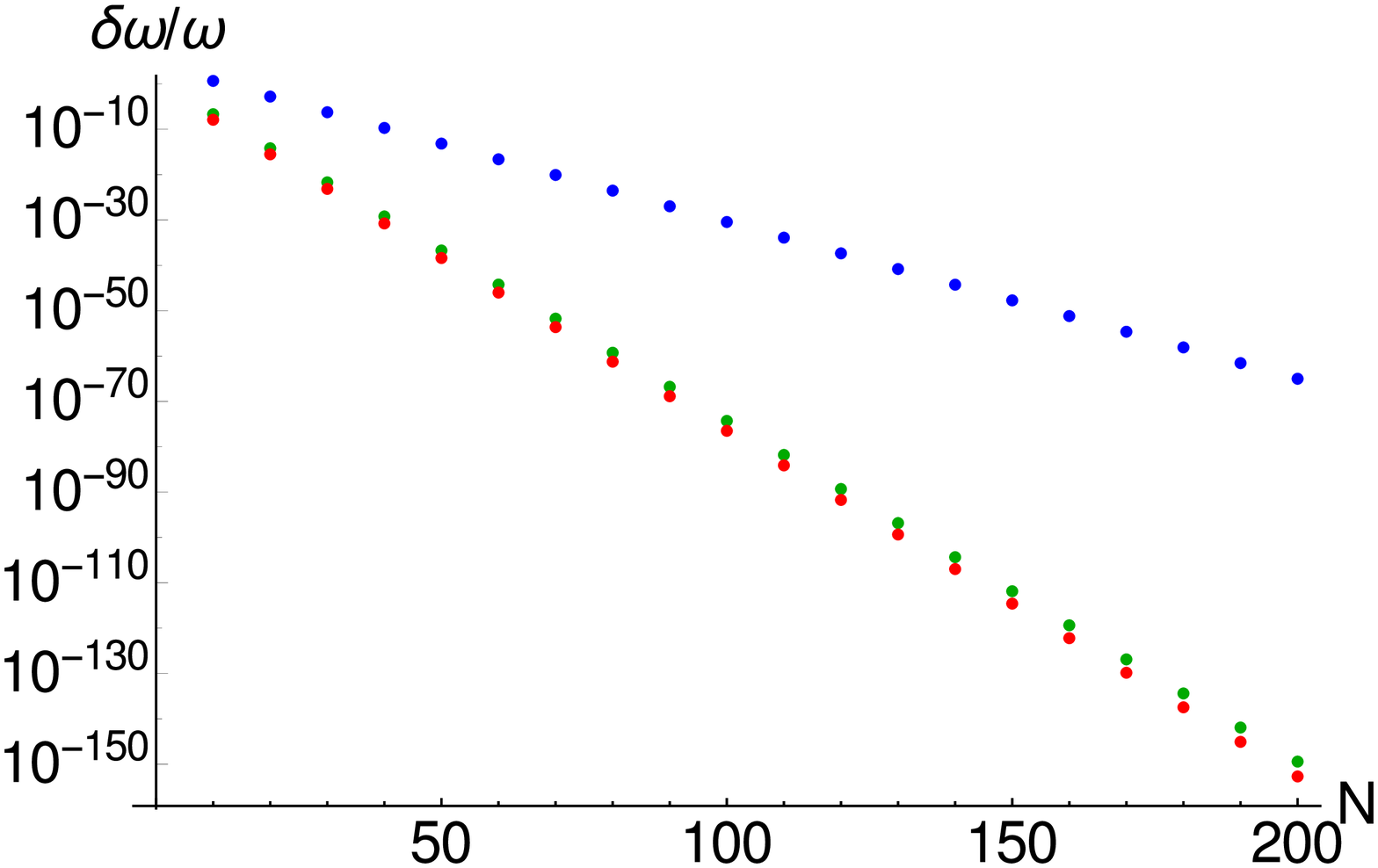}\includegraphics[height=5cm]{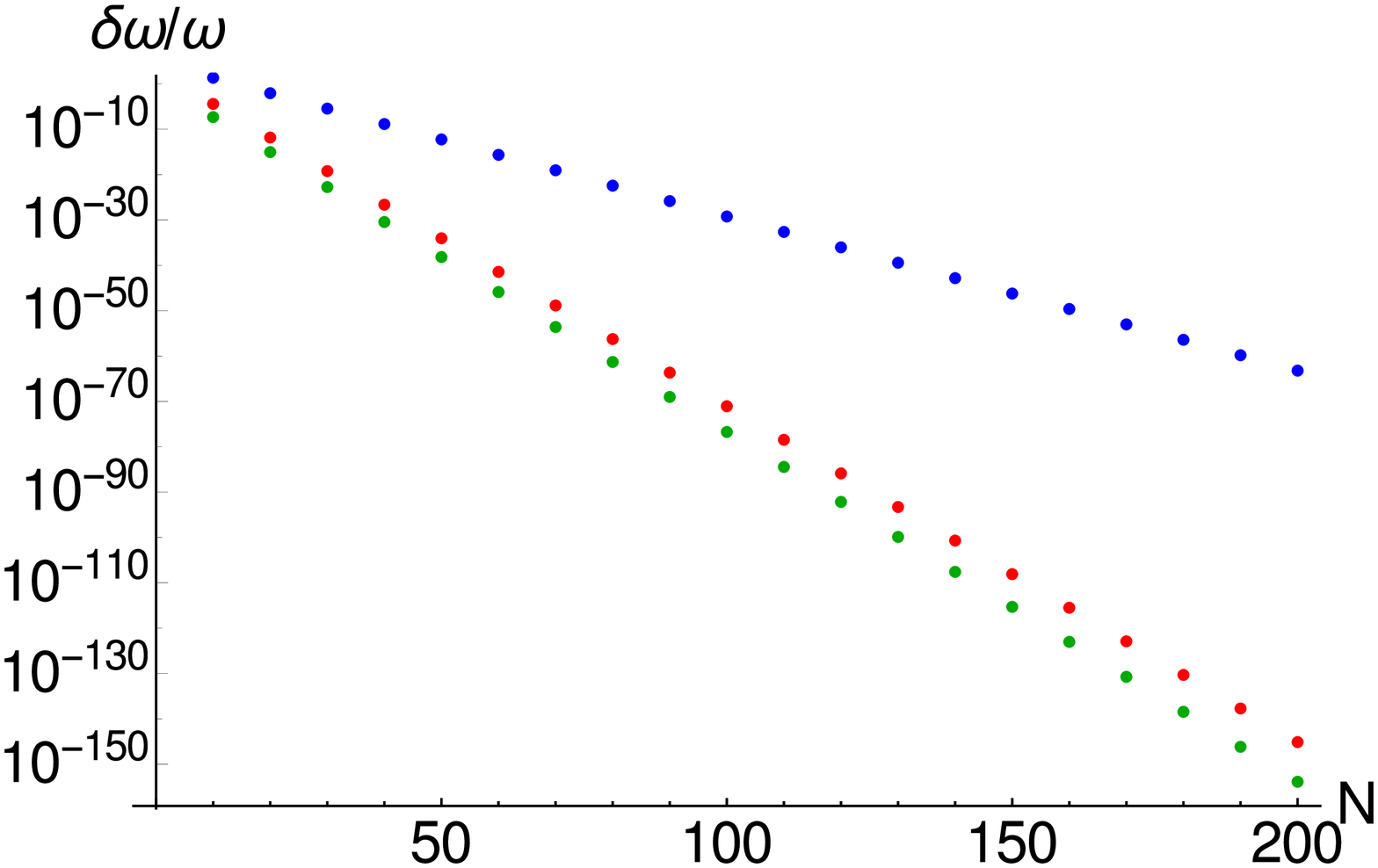}
		\caption{The accuracy of the first three numerical QNM frequencies from the first set in comparison with analytical formula for $M=0.5$ for $\mathcal{Z}_1$ (left panel) and $\mathcal{Z}_2$ (right panel).
			The horizontal axis shows the number of grid points along the radial coordinate and green, red and blue dots correspond to the  $n=0, 1, 2$ QNMs respectively. It is interesing that in this case the first mode in the $\mathcal{Z}_1$ family has higher accuracy than the zeroth mode.}  
		\label{QNMaccuracy}
	\end{center}	
\end{figure}

Therefore, while the $\mathcal{Z}_1$ channel is stable for $M>1$,  the $\mathcal{Z}_2$ channel is always unstable for any value of the parameter $M$, in the regime of our interest. 
Note that this is true for any 3D gravity theory which leads to the linearized equation in TT-gauge, \eqref{2ordereq} with a massive mode. 
This is a fascinating result which we found by implementing the numerical analysis for the first time.
In \cref{QNMaccuracy} we show the accuracy of our numerical results comparing with the analytical QNMs given in \cref{QNM-analytic-k1,QNM-analytic-k} for the first three modes in the first set.

\begin{figure}[ht]
	\begin{center}
		\includegraphics[height=4.5cm]{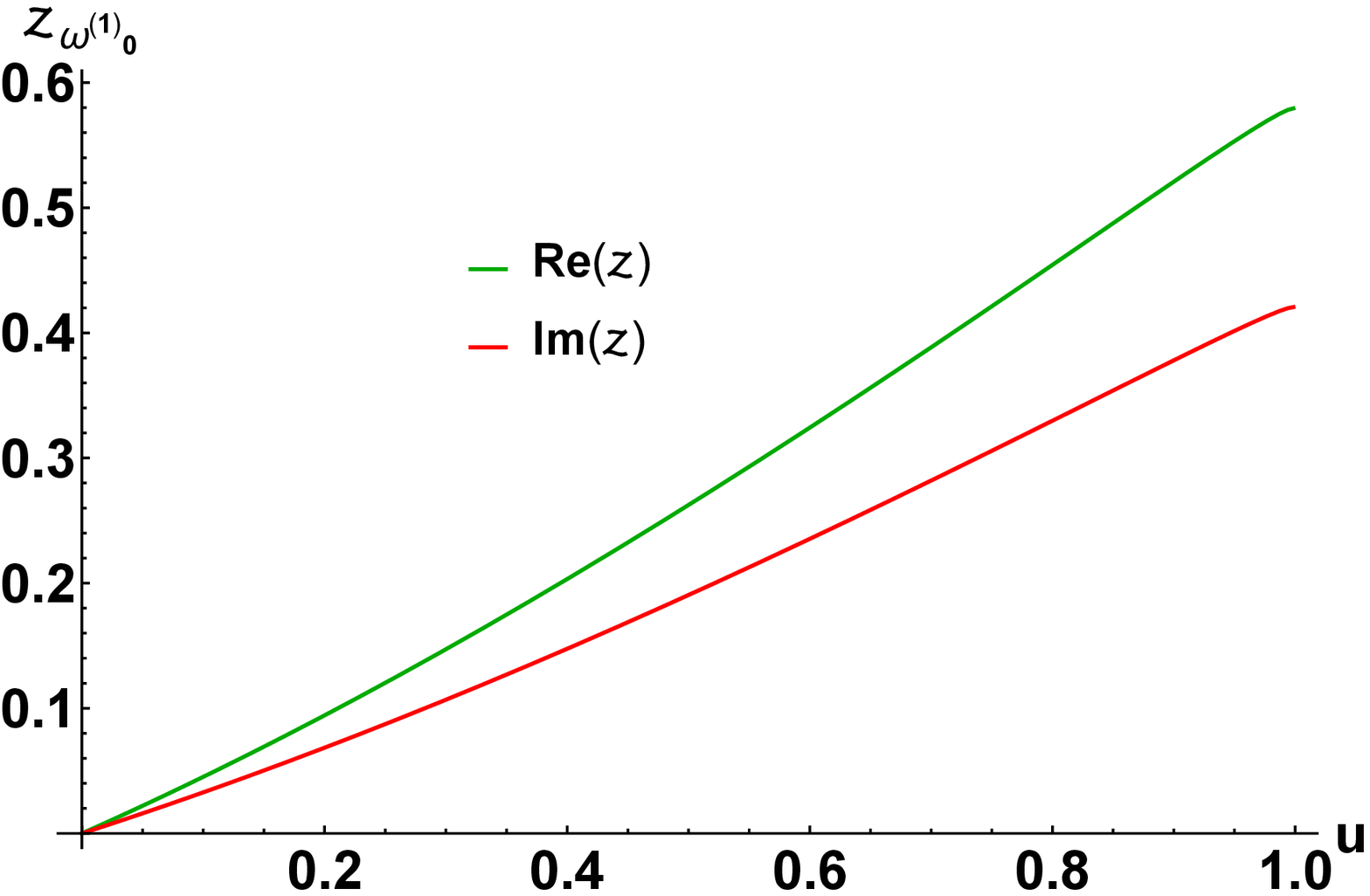} \hspace{20pt}
		\includegraphics[height=4.5cm]{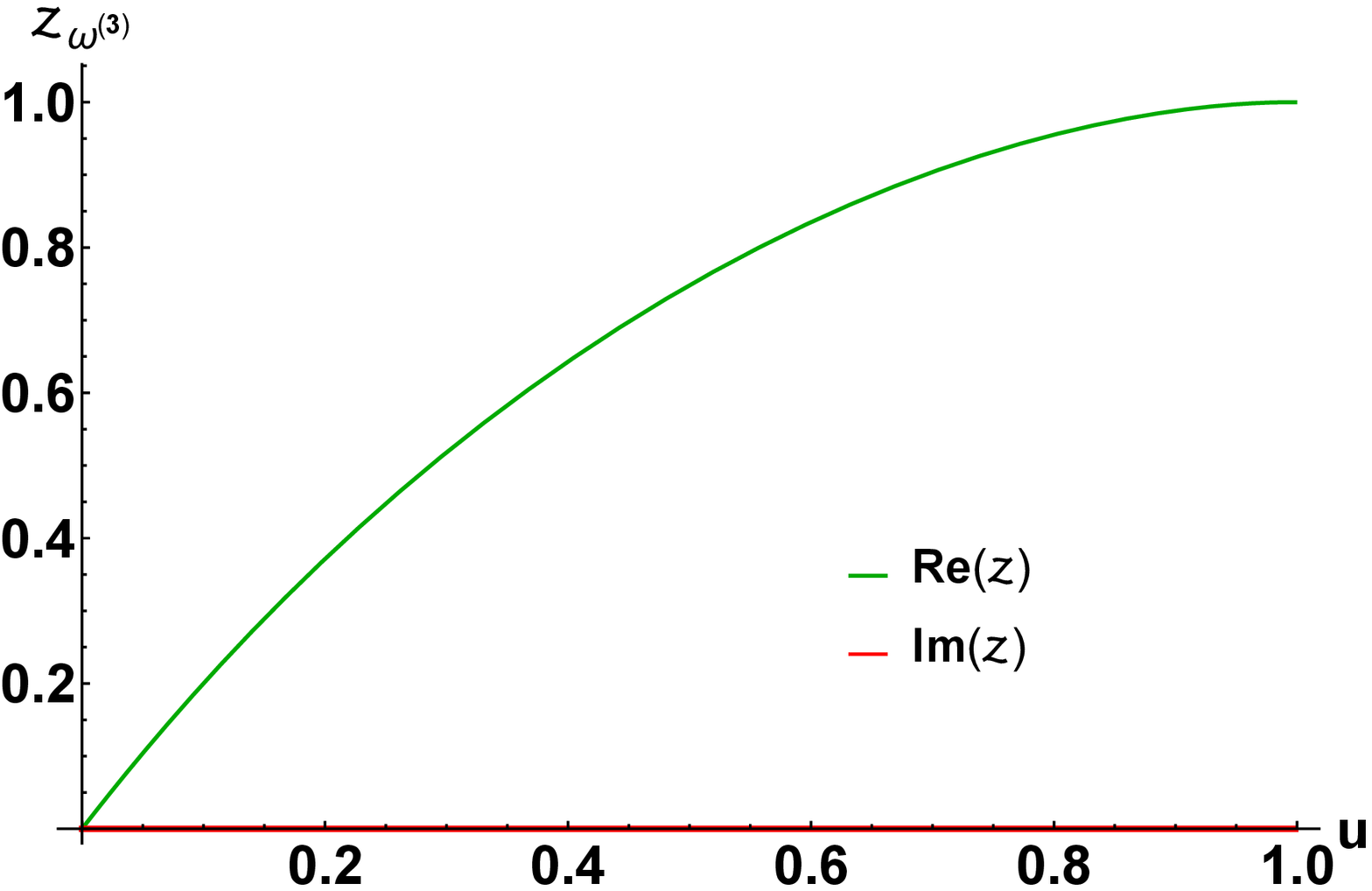}
		\includegraphics[height=4.5cm]{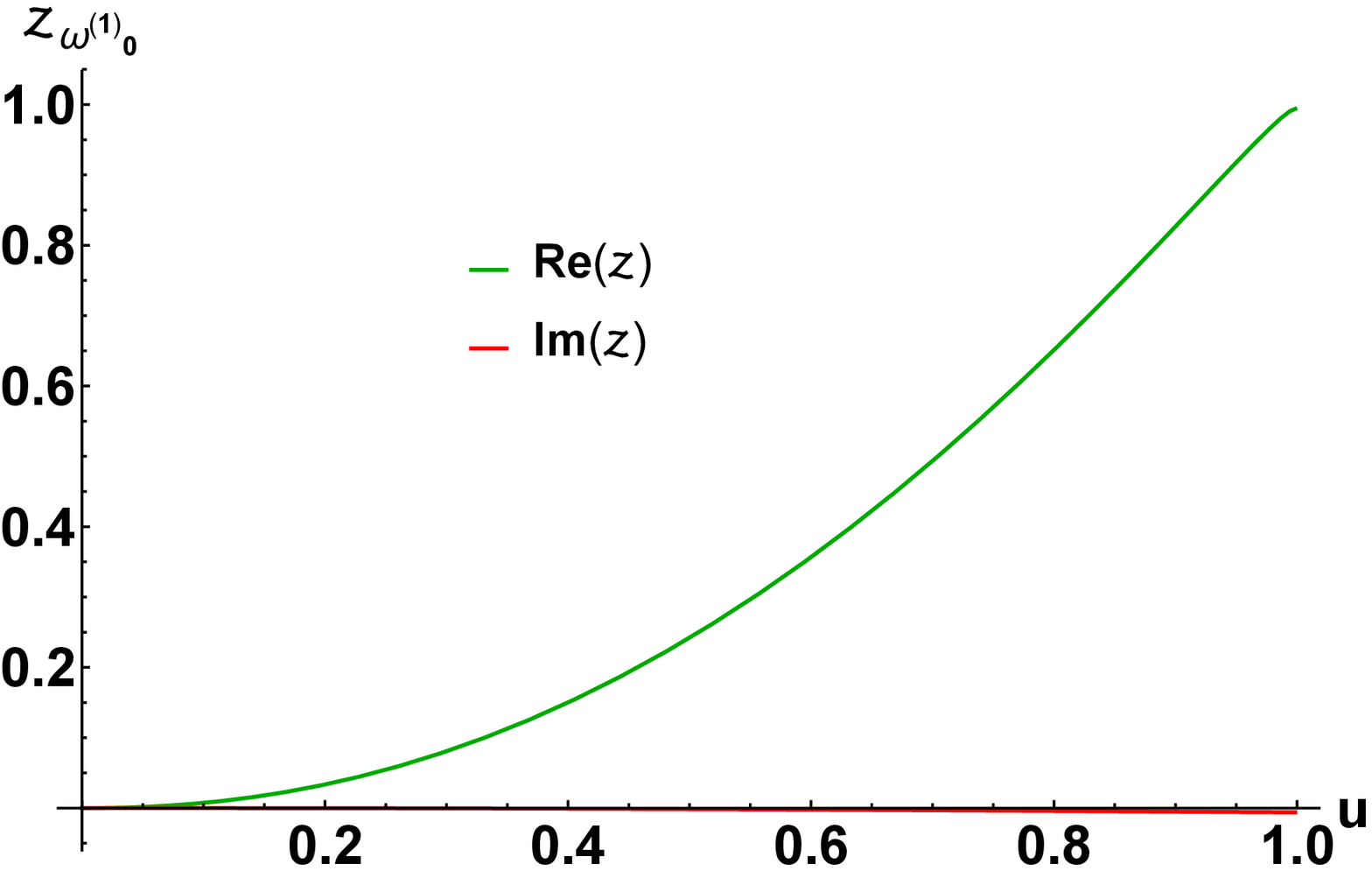}\hspace{20pt}
		\includegraphics[height=4.5cm]{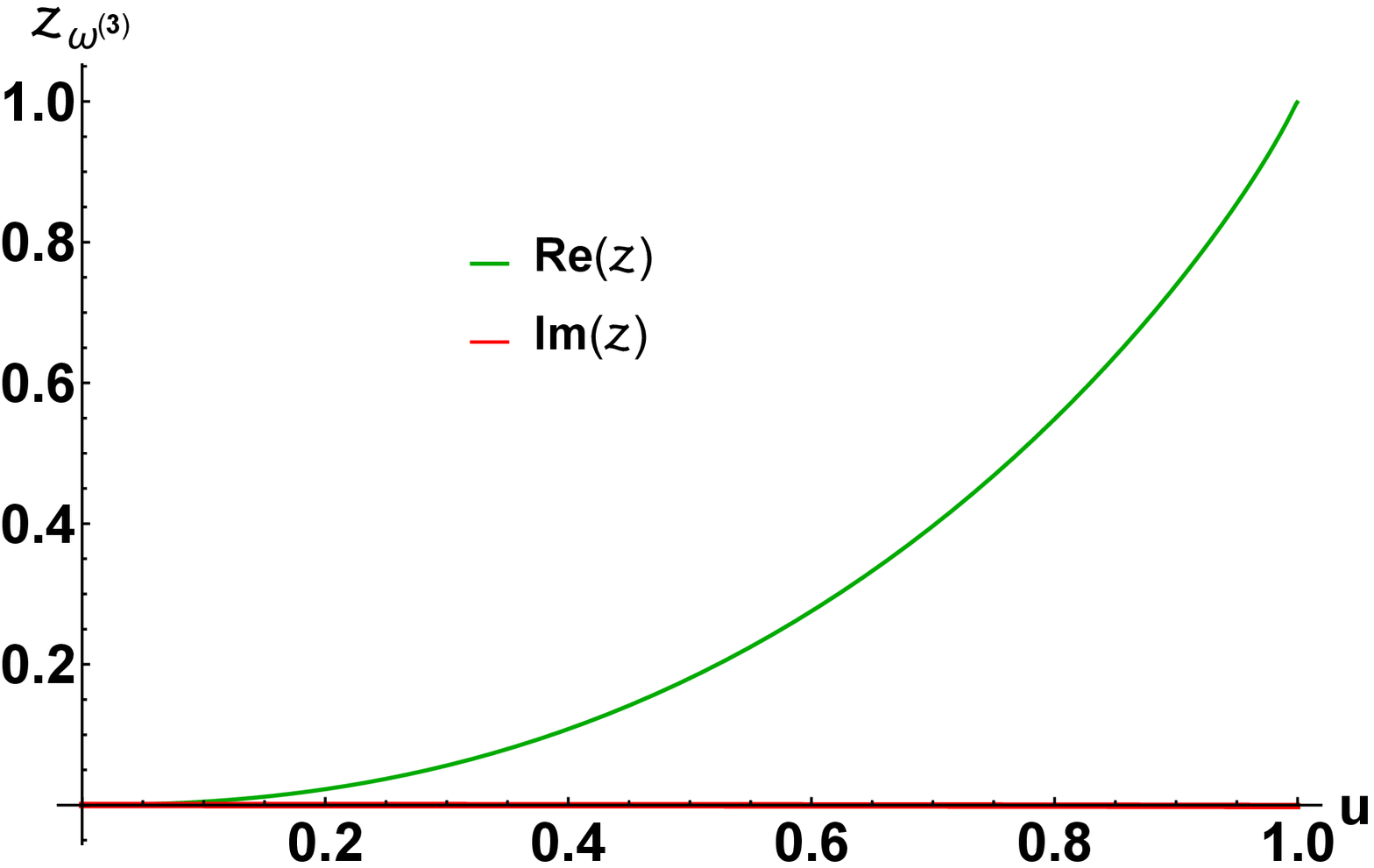}
		\caption{The eigenmodes corresponding to the first two QNMs with smallest absolute value of the frequencies at zero angular momentum are plotted for $M=0.5$ (upper panels) and $M=1.1$ lower panels.
			The left panels are the zeroth mode in the first set and the right panels are the single mode in the third set.}  
		\label{eigenvectors}
	\end{center}	
\end{figure}

By looking at the Eigen-vectors of the QNMs, we can justify our results and show that there is no pathology in the new modes that we found. 
In  \cref{eigenvectors} we show the real and imaginary parts of the $\mathcal{Z}_2$ Eigen-vectors for the zeroth mode of the first set of QNMs (left panels) and the single mode  in the third set of QNMs (right panels) at zero angular momentum for $M=0.5$ (upper panels) and $M=1.1$ (lower panels). 
According to our results given in \cref{qnmk}, for $M=0.5$ the unstable mode is  $\omega^{(1)}_0$ but for  $M=1.1$ the instability is due to the $\omega^{(3)}$. 
The expected behaviour of the Eigen-vectors at the horizon (regularity condition) and at the asymptotic region ($\mathcal{Z}\sim u^{2M}$) could strongly support our claim about instability of BTZ solutions in parity-even 3D-gravity models.

It is very interesting that one can check analytically if the third set frequency corresponds to true QNM eigenmode.
It is straightforward to see that plugging  $\omega=-i\,\left(1-M^2\right)$ in  \cref{Z2-eq} reduces 
to\footnote{We would like to thank our referee for his suggestion to  add  this analytical analysis which improves the manuscript and support our numerical results.}
\begin{equation}\label{3rdmode}
\mathcal{Z}''(u)+\tfrac{\left(2 M^2 u-2 M (u^2-1)+(u-2) u-1\right) }{u (u^2-1)}Z'(u)+\tfrac{(M^2-1) \left(-2 M (u+1)+u^2+1\right) }{(u-1) u (u+1)^2}Z(u)=0~,
\end{equation}
which has analytical solution in terms of hypergeometric functions as
\begin{align}
\mathcal{Z}(u)=&\mathcal{A}\,(u+1)^{M^2-1} \, _2F_1\left(\tfrac{1}{2}M (M-1),\tfrac{1}{2} \left(M^2-M-2\right);1-M;u^2\right)\nn\\
+&\mathcal{B}\,u^{2 M} (u+1)^{M^2-1} \, _2F_1\left(\tfrac{1}{2} M (M+1),\tfrac{1}{2} \left(M^2+M-2\right);1+M;u^2\right)~.
\end{align}
As mentioned earlier, QNM Eigen-mode  is defined with source free boundary condition at the asymptotic boundary which means $\mathcal{A}=0$.
Also since the \cref{3rdmode} is a linear equation,  one may rescale the Eigen-mode and the QNM mode is given by,
\begin{equation}
\mathcal{Z}(u)=u^{2 M} (u+1)^{M^2-1} \, _2F_1\left(\tfrac{M (M+1)}{2} ,\tfrac{(M^2+M-2)}{2} ;M+1;u^2\right),\qquad \omega=-i\,\left(1-M^2\right).
\end{equation}
We would like to point that one can find the Eigen-modes of  other QNM frequencies in the first and second channels by plugging   \cref{QNM-analytic-k1,QNM-analytic-k} into  linearized \cref{Z1-eq,Z2-eq}. The results for couple of the lowest modes of the first set in the first branch are
\begin{align}
&\mathcal{Z}(u)=u^{2 M} (u+1)^{-M},& \omega^{1,1}_0=-i(M-1),\\
&\mathcal{Z}(u)=u^{2 M} (u+1)^{-M-2} \left((u+1)^2-M\right),& \omega^{1,1}_1=-i(M+1),\\
&\mathcal{Z}(u)=u^{2 M} (u+1)^{-M-4} \left(\tfrac{1-M}{4}+u\right),& \omega^{1,1}_2=-i(M+3),\\
&\mathcal{Z}(u)=u^{2 M} (u+1)^{-M-6} \left(\tfrac{1-M^2}{12}+\tfrac{1+M}{2} u-\tfrac{1+M}{4} u^2+u^3\right),& \omega^{1,1}_3=-i(M+5),
\end{align}
which are fascinating results  supporting our numerical analysis. 
While we found these expressions one by one, it might be possible to find a general analytical solution for  $n$th frequency.

\subsection{$k\neq0$}

Now let's turn to the finite angular momentum, $k\neq0$.
In this case, by using the TT-gauge constraints, one can show that the linearized equation \eqref{2ordereq} will reduce to two coupled equations for the metric perturbations $h_{r\phi}$ and $h_{\phi\phi}$. 
Hiring the same redefinitions introduced in \cref{redefine1,redefine2}  one can find the  linearized equations in new functions which  are too lengthy and  given in appendix \ref{knotzero}.
Similar to the $k=0$ case, the only free parameter in the equations is $M$ and there is no explicit temperature dependency. But at finite angular momentum the equations are coupled. 
The near boundary analysis shows the same expansion of the $\mathcal{Z}_i$, given in \cref{near-boundary}.
Again, our numerical results are as accurate such that the following analytical forms of the frequencies, which are  classified in three sets, can be proposed
\begin{align}\label{qnmk}
&\omega^{(1)}_n(k)=-i\,\left(M-1+2\,n\right)\pm k~,\qquad \omega^{(2)}_n=-i\,\left(M+3+2\,n\right)\pm k~, \qquad n=0, 1, 2, \cdots~,\nonumber\\
&\omega^{(3)}=-i\,\left(1-M^2+\left(\pm k\right)^2\right)~.
\end{align}
The third set has completely different dependency  not only in the parameter $M$ but also in angular momentum $k$ such that they are purely imaginary for any value of the parameters in the domain of our interest. Note that there is a degeneracy for any value of angular momentum $k$ and to our knowledge, this is the first example in asymptotically AdS black holes that there are unique modes which do not belong to any tower of the QNMs.

On the other hand, for any value of parameter $M$ and finite angular momentum $k\neq0$, there are two unstable modes either in the first set or in the third set of the frequencies given in \cref{qnmk}. 
In this case all the metric fluctuations are coupled which means the static BTZ black hole for any theory, 
which has massive mode satisfying the general equation \eqref{2ordereq}, 
is dynamically unstable with respect to any non zero angular momentum perturbation.

While the structure of the first two sets is similar to what we found  from analytical studies  in \cref{QNMs}, there is  an obvious difference in the real part of the modes.
Our numerical modes can move in both left and right directions since they have both signs in their real part. This is the second new results we found using the numerical approach to solve the second order linearized equation \eqref{2ordereq}.
Unfortunately, in contrast to the previous subsection, we could not find the analytical Eigen-modes of linearized coupled equations at non-zero angular momentum.
But the   numerical results are as accurate as the zero angular momentum case.

\section{Discussion and Outlook}\label{discussion}

The BI theory of gravity is interesting for different reasons. 
It could be considered as a re-summation of infinite number of higher derivative corrections to the Einstein-Hilbert action while by expanding the BI action in parameter $1/m^2$, 
one may get NMG theory or its extended versions to any order. 
Therefore, it could provide a proper setup to investigate the effects of higher derivative corrections to the physical properties that one would like to study.

In this paper we found regimes of the parameters of the BI theory in which static BTZ black holes are solutions to the equations of motion.
Then we investigate the quasinormal modes of these solutions to see if there is a range of parameters which leads to stable black holes. 

We use the known analytical approach \cite{Li:2008dq} to decompose the second order differential equation \eqref{2ordereq}, which specifies the quasinormal modes, to the first order differential equations \eqref{1ordereq}, 
and the  results for left/right moving modes are given in \cref{QNMs}.
While the solutions to the later equations clearly satisfy the former one, there is no guarantee that all the solutions to the original equation could be found in this way. 
Therefore, we use the numerical tools to tackle the equation \eqref{2ordereq} with ingoing boundary condition at the horizon and source free boundary condition at the asymptotic region.

Our numerical analysis is based on a general form of the linearized equation \eqref{2ordereq}  in TT-gauge which let us to make some comments on any 3D parity even massive gravity theories including NMG \cite{Bergshoeff:2009hq}, its extensions \cite{Sinha:2010ai,Afshar:2014ffa} and ZDG \cite{Bergshoeff:2013xma , Bergshoeff:2014bia}.
On the other hand, the numerical results are accurate which let us to propose analytic expression for the QNM frequencies given in \cref{QNM-analytic-k},
and they are  promising to have analytical solutions for the second order linearized equation. 
By plugging  the QNM frequencies \cref{QNM-analytic-k1,QNM-analytic-k} in to the \cref{Z1-eq,Z2-eq} we found analytical solutions for some modes in vanishing angular momentum which confirms our numerical results.
Finding a general form of the analytical eingenmodes could be an interesting open question which needs further investigations.
We didn't study the quasinormal frequencies of the solutions at the critical point at which Logarithmic modes are new solutions to the equations. 
This has been studied from different point of views in various cases in  \cite{Grumiller:2008qz,Alkac:2017vgg,Kim:2012rz,Grumiller:2009sn}.

As we know, BTZ black hole is a solution to the three dimensional Einstein-Hilbert and its massive extensions, e.g. TMG, NMG and BI-model.
There is a common feature in different dimensions that the solutions to the Einstein-Hilbert action are solutions to its extended massive gravity theories as well.
The stability of these solutions in massive gravity models have been studied in various cases. 
For example, in NMG theory it was shown that the warped AdS geometry is unstable in all range of parameters \cite{Ghodsi:2012fg} while
in four dimensional gravity theories instability of Schwarzschild black holes and Kerr geometries have been studied in \cite{Babichev:2013una , Brito:2013wya}.
Therefore this kind of instability may have the same origin and investigating this point could help us to understand the physics of massive spin-2 excitations.

What we found shows two sets of  quasinormal frequencies, with a structure similar to the analytical results, and  new unique modes which do not belong to any tower of the QNMs in asymptotically AdS backgrounds. 
On top of that, the dependency of the modes on the only parameter of the linearized equations is such that it leads to unstable quasinormal modes in any regime of the parameters  in any 3D massive theory which has a linearized equation given in \cref{2ordereq}. 

Although we show instability of BTZ black holes in parity-even massive theories, we did not study what it decays to.
The following steps has to be taken to answer this question.
It is known that solutions to three dimensional gravity theories with higher derivative terms are richer than Einstein gravity and have not been completely classified \cite{Ghodsi:2010ev,Gurses:2011fv,Gurses:2012db,Altas:2015dfa}.
Therefore as the first step, one has to find all the  solutions (including time dependent geometries) of the equations of motion of the corresponding model.
Then, one may study the full time evolution of the perturbed BTZ within the theory.
This is a very interesting open question which is beyond the scope of this paper and needs further studies.
%
%

Last but not least, we would like to compare our results with the QNMs of  5D gravity theories with negative cosmological constant and  at the presence of the higher derivative curvature terms in the action studied in \cite{Grozdanov:2016fkt , Grozdanov:2018gfx}. 
It was shown that at large but finite coupling, the branches of the QNMs become more dense and lift up towards the real axis, which may lead to forming a branch cut in the limit of zero coupling.
In contrast, our analysis reveals a different structure either for the full BI-action or for its truncated expansion at any order in $1/m^2$,  in which the QNMs have finite distances in the frequency complex plane and there is no evidence of forming any branch cut. Comparing these two structures in three and five dimensional theory of gravity in asymptotically AdS geometries needs more investigations. 

\subsubsection*{\bf Acknowledgments}

We would like to thank R. Janik and B. Tekin  for their comments on the first version of this manuscript.
We would also like to thank H. R. Afshar and G. Alkac for discussion on different aspects of 3D-gravity.
H.S. would like to thank Jagiellonian University, where part of this research was done. 
The abstract tensor calculations in this paper has been carried out by Mathematica package xAct \cite{Nutma:2013zea,Brizuela:2008ra}.

\appendix

\section{Stress tensor for BI gravity}\label{Btensor}

In order to find the equation of motion \eqref{EOM} one has to apply integration by part.  
The total derivative of this integration leads to a boundary term which has the following well-known form

\be 
\int_{\pd\mathcal{M}}\sqrt{-h}\ n_{\alpha}J^{\alpha},
\ee
 where the expression for $ J^\alpha $ is given by,
\begin{align}\label{jaryan}
J_{\alpha}&=\sigma\big[- \delta \mathit{g}^{\gamma}{}_{\gamma} \nabla_{\alpha}\mathcal{V}^{\beta}{}_{\beta} + \delta \mathit{g}_{\beta \gamma} \nabla_{\alpha}\mathcal{V}^{\beta \gamma} -  \mathcal{V}^{\beta \gamma} \nabla_{\alpha}\delta \mathit{g}_{\beta \gamma} + \mathcal{V}^{\beta}{}_{\beta} \nabla_{\alpha}\delta \mathit{g}^{\gamma}{}_{\gamma} + \delta \mathit{g}^{\gamma}{}_{\gamma} \nabla_{\beta}\mathcal{V}_{\alpha}{}^{\beta} + \mathcal{V}^{\beta \gamma} \nabla_{\beta}\delta \mathit{g}_{\alpha \gamma} \nonumber\\&-  \mathcal{V}^{\beta}{}_{\alpha} \nabla_{\beta}\delta \mathit{g}^{\gamma}{}_{\gamma} + \mathcal{V}^{\beta \gamma} \nabla_{\gamma}\delta \mathit{g}_{\alpha \beta} -  \mathcal{V}^{\beta}{}_{\beta} \nabla_{\gamma}\delta \mathit{g}_{\alpha}{}^{\gamma} -  \delta \mathit{g}_{\beta \gamma} \nabla^{\gamma}\mathcal{V}_{\alpha}{}^{\beta} -  \delta \mathit{g}_{\beta \gamma} \nabla^{\gamma}\mathcal{V}^{\beta}{}_{\alpha} + \delta \mathit{g}_{\alpha \gamma} \nabla^{\gamma}\mathcal{V}^{\beta}{}_{\beta}\big]~.
\end{align}
Although the expression seems too complicated and finding the proper boundary terms for general metric background is impossible, for a maximally symmetric spacetime background \cref{jaryan} reduces to
\begin{equation}\label{BT2}
J_{a}=\lambda \sigma (\nabla_{\alpha}\delta \mathit{g}^{\beta}{}_{\beta} - \nabla_{\beta}\delta \mathit{g}_{\alpha}{}^{\beta})
\end{equation}
The terms in the pare$n$thesis of this surface integral is nothing but what we get from variation of the Einstein-Hilbert action and the pre-factor $\sigma\lambda$  is the fingerprint of the BI theory.
It is well-known that using these terms one can obtain the Gibbons-Hawking-York  boundary term and the Brown-York stress  tensor, (see \emph{e.g.} \cite{Padmanabhan:2014lwa} ). 
Therefore, the boundary terms and stress  tensor for the three dimensional BI theory of gravity on maximally symmetric backgrounds is a multiplication of the  Einstein-Hilbert ones with $\sigma\lambda$, namely
\bea
\mathcal{L}_{\text{boundary}}=2\frac{\sigma\lambda}{\kappa_3^2}\int d^2x \sqrt{|h|} K~,\qquad T^{\mu\nu}=\sigma\lambda (K^{\mu\nu}-h^{\mu\nu}K)~. 
\eea

To compute the free energy of a solution one should evaluate the full on-shell action which is 
\begin{equation}\label{lag}
\mathcal{L}_{T}=\mathcal{L}_{\text{BI}}+\mathcal{L}_{\text{boundary}}+\mathcal{L}_{\text{ct}}~,
\end{equation}  
where $\mathcal{L}_{\text{ct}}$ is the boundary local counterterm  \cite{Balasubramanian:1999re}.
By comparing the total action \cref{lag} with the three dimensional Einstein gravity theory with negative cosmological constant, 
it is easy to show that, for a locally AdS geometry, the free energy has the same pre-factor structure as  we found for the central charge and the boundary stress tensor given in \cref{ccharge2,stress-tensor} respectively.

\section{Non-vanishing angular momenta}\label{knotzero}

The coupled linearized equations with nonzero angular momentum are more involved. 
In this appendix we show the explicit form of these equations for completeness.
As we expected, at finite angular momentum, these are coupled equations given by
\bea
&&\alpha_0\, \mathcal{Z}_1(u)+\alpha_1\, \mathcal{Z}'_1(u)+\alpha_2\, \mathcal{Z}''_1(u)+\beta_0\, \mathcal{Z}_2(u)+\beta_1\, \mathcal{Z}'_2(u)=0~,\\
&&\widetilde{\alpha}_0\, \mathcal{Z}_1(u)+\widetilde{\alpha}_1\, \mathcal{Z}'_1(u)+\widetilde{\beta}_0\, \mathcal{Z}_2(u)+\widetilde{\beta}_1\, \mathcal{Z}'_2(u)+\widetilde{\beta}_2\, \mathcal{Z}''_2(u)=0~,
\eea
where

{\tiny
\begin{align}
&\alpha_0=u \left(k^4 u^2+k^2 (M-1)^2 \left(u^2+1\right)+(M-3) (M-1)^2 (M+1)\right) \left(M^2 \left(\left(2 k^2-1\right) u^2-1\right)+\left(k^2+1\right) u^2 \left(k^2 u^2+1\right)+M^4\right)\nn\\
&+i   \bigg(2 k^6 (M-1) u^6+k^4 u^4 \left(6 M^3-11 M^2+(M-2) (M+1) u^2-M+6\right)+k^2 \left(3 (M-1) M \left(2 (M-1) M^2+5\right) u^2-2 \left(M^4+M^3+4 M-3\right) u^4+(M-3) M u^6\right)\nn\\
&+M \left(M^2-1\right) (M-u) (M+u) \left(2 M^2-3 M \left(u^2+1\right)+9 u^2+1\right)\bigg)\,\omega-u \, \bigg(u^2 \left(k^2 \left(M \left(-4 M^2+M+4\right)+2\right)-(M-1) (M+1) ((M-10) M+6)\right)\nn\\ 
&+k^2 u^4 \left(2 k^2 (M+1)+M (3 M-4)+2\right)+3 M^2 \left(M \left(-2 M^2+M+2\right)-1\right)\bigg)\,\omega^2-i  \left(6 k^2 (M-1) u^4+M u^2 (-2 M^2+3 (M-3) u^2+3 M-1)\right) \omega ^3+3 (2 M-1) u^3 \omega ^4,\nn\\ \nn \\
&\alpha_1=\left(u^2 \left(2 \left(k^2+1\right) M+k^2-2 M^3+5 M^2-5\right)-k^2 (2 M-3) u^4+2 M^3-M^2-2 M+1\right) \left(M^2 \left(\left(2 k^2-1\right) u^2-1\right)+\left(k^2+1\right) u^2 \left(k^2 u^2+1\right)+M^4\right)\nn\\
&-2 i \omega  \bigg(k^2 \left(k^2+1\right) u^7 \left(k^2+M-2\right)+u^3 \left(k^2 (M (M (M (3 M+2)+2)-5)-5)-(M-1)^2 (M+1) (M (3 M-2)+1)\right)\nn\\
&+u^5 \left(k^4 M (3 M-1)-2 k^2 \left(M^3-2 M+3\right)+3 (M-2) \left(M^2-1\right)\right)+M^2 \left(M^2-1\right) (M (M+3)-1) u\bigg)+u^2 \omega ^2 \bigg(u^2 \left(k^2 (7-2 M (2 M+3))-\left(M^2-1\right) (2 M-11)\right)\nn\\
&+k^2 u^4 \left(2 k^2+6 M-7\right)-6 M^4+2 M^3+5 M^2-2 M+1\bigg)+2 i u^3 \omega ^3 \left(3 u^2 \left(k^2+M-2\right)-M^2-3 M+1\right)-6 u^4 \omega ^4,\nn\\ \nn \\
&\alpha_2=\left(u^2-1\right) u \left(3 k^4 M^2 u^4+\left(M^2-1\right) u^2 \left(\left(3 k^2-1\right) M^2+k^2+1\right)+\left(k^6+k^4\right) u^6+M^2 \left(M^2-1\right)^2\right)+i \left(u^2-1\right) u^2 \omega  \bigg(M^2 \left(\left(3-2 k^2\right) u^2+3\right)\nn\\
&+u^2 \left(k^2 \left(\left(k^2+1\right) u^2+5\right)-3\right)-3 M^4\bigg)-\left(u^2-1\right) u^3 \omega ^2 \left(3 k^2 u^2-M^2+1\right)-3 i \left(u^2-1\right) u^4 \omega ^3,\nn\\ \nn 
\end{align}
\begin{align}
&\beta_0=-i k \bigg((M-1) u^2 \left(k^2 (2 M (M (3 M+2)-3)-1)+M \left(5 M \left(M^2-2\right)-6\right)-1\right)+k^2 \left(k^2+1\right) u^6 \left(2 k^2+M+1\right)+(M-1)^2 M^2 \left(2 M^2+M-1\right)\nn\\
&+u^4 \left(k^4 \left(6 M^2+M-1\right)+k^2 (2 M (3 M (M+1)-1)-4)+3 (M-1) (M+1)^2\right)\bigg)-2 k (M-1) u \omega  \left(M^2 \left(\left(2 k^2+3\right) u^2-1\right)+u^2 \left(\left(k^2+1\right) k^2 u^2+k^2-3\right)+M^4\right)\nn\\
&+i k u^2 \omega ^2 \left(3 u^2 \left(2 k^2+M+1\right)+8 M^3-2 M^2-5 M-1\right)+6 k (M-1) u^3 \omega ^3,\nn\\ \nn \\
&\beta_1=i k u \left(M^2 \left(3 u^2-1\right) \left(\left(2 k^2+1\right) u^2-3\right)+\left(k^2+1\right) u^2 \left(u^2 \left(k^2 \left(u^2+1\right)-3\right)+5\right)+M^4 \left(5 u^2-3\right)\right)\nn\\
&+2 k u^2 \omega  \left(M^2 \left(\left(2 k^2+3\right) u^2-1\right)+u^2 \left(\left(k^2+1\right) k^2 u^2+k^2-3\right)+M^4\right)-i k u^3 \omega ^2 \left(8 M^2+3 u^2-5\right)-6 k u^4 \omega ^3 \nn 
\end{align}
\begin{align}
&\widetilde{\alpha}_0=4 i \omega ^2 \left(2 k^5 u^6+k^3 u^4 \left(4 M^2+(M+2) u^2-M-6\right)+k M u^2 \left(2 M^3-M^2+((M-2) M-2) u^2-2 M+u^4+1\right)\right)-8 k u^3 \omega ^3 \left(\left(k^2+1\right) u^2+M^2-1\right),\nn\\
&+4 k u \omega  \left((M^2-1) u^2 \left(M (-2 k^2+(M-4) M+2)-2\right)+k^2 (k^2+1) M u^6-u^4 \left(k^4 (M+2)+k^2 (M (3-2 (M-2) M)-2)+(M-2) (M^2+1)\right)-(M-1)^3 M (M+1)\right)\nn\\ \nn \\
&\widetilde{\alpha}_1=-4 k \left(u^2-1\right) u^2 \omega  \left(\left(2 k^2-1\right) \left(M^2-1\right) u^2+\left(k^4+k^2\right) u^4+M^4-1\right)-4 i k \left(u^2-1\right) u^3 \omega ^2 \left(\left(k^2+1\right) u^2+M^2-1\right),\nn\\ \nn \\
&\widetilde{\beta}_0=u \left(k^2+(M+1)^2\right) \left(k^2 u^2+M^2-1\right) \left(u^2 \left(k^2 (2 (M-2) M-1)-(M-1)^2\right)+k^2 \left(k^2+1\right) u^4+(M-2) (M-1)^2 M\right)+i \omega  \bigg(M^2 (2 M-1) \left(M^2-1\right)^2\nn\\
&+ (M-1) u^2 \left(k^2 \left(6 M^4+8 M^3+M-3\right)+(1-3 M) \left(M^2-1\right)^2\right)+k^2 \left(k^2+1\right) u^6 \left(2 k^2 (M+4)+(M+1)^2\right)+3u^4 (M^2-1)^2\nn\\
&+u^4 \left(k^4 (M (M (6 M+11)-4)-7)-2 k^2 (M ( M (M^2-4M-5)+6)+8)\right)\bigg)+\omega ^2 \bigg( M (M^2-1) (2 M+1) (3 M-2) u-k^2 u^5 \left(2 k^2 (M+8)+M (3 M+2)+4\right)\nn\\
&+u^3 \left(k^2 (M (M (4 M-13)-4)+4)+(M^2-1) ((M-6) M-4)\right)\bigg)+i u^2 \omega ^3 \left((M-1)^2 (2 M-1)-3 u^2 \left(2 k^2 (M+2)+M^2-1\right)\right)+3 (2 M+1) u^3 \omega ^4,\nn 
\end{align}
\begin{align}
&\widetilde{\beta}_1=(M^2-1) u^2 \left(k^2 \left(6 M^3+M^2+2 M+1\right)-2 M^5+3 M^4-4 M^2+2 M+1\right)+k^2 u^6 \left(k^4 (2 M+3)+k^2 \left(-6 M^3+M^2+2 M+5\right)-2 M^2+2\right)\nn\\
&+u^4 \left(k^4 \left(6 M^3+5 M^2-2\right)-k^2 (M-1)^2 \left(6 M^3+7 M^2+4 M+3\right)+(2 M-1) \left(M^2-1\right)^2\right)-k^4 \left(k^2+1\right) (2 M+1) u^8+M^2 (2 M-1) \left(M^2-1\right)^2\nn\\
&-i \omega  \bigg(u^3 \left(k^2 \left(6 M^4+4 M^3-10 M^2-10 M+1\right)-6 M^5+7 M^4-8 M^2+6 M+1\right)+k^2 \left(k^2+1\right) u^7 \left(2 k^2+2 M+1\right)+M^2 \left(2 M^4+6 M^3-7 M^2-6 M+5\right) u\nn\\
&+u^5 \left(k^4 \left(6 M^2-2 M-3\right)-2 k^2 \left(2 M^3-5 M^2-4 M+8\right)+6 M^3-3 M^2-6 M+3\right)\bigg)+u^2 \omega ^2 \bigg(u^2 \left(k^2 \left(-4 M^2-6 M+7\right)-2 M^3+7 M^2+2 M-7\right)\nn\\
&+k^2 u^4 \left(2 k^2+6 M-1\right)-6 M^4+2 M^3+7 M^2-2 M-1\bigg)+i u^3 \omega ^3 \left(u^2 \left(6 k^2+6 M-3\right)-2 M^2-6 M-1\right)-6 u^4 \omega ^4,\nn \\ \nn \\
&\widetilde{\beta}_2=u \left(u^2-1\right) \left(3 k^4 M^2 u^4+\left(M^2-1\right) u^2 \left(k^2 \left(3 M^2+1\right)-M^2+1\right)+\left(k^6+k^4\right) u^6+M^2 \left(M^2-1\right)^2\right)\nn\\
&+i u^2 \left(u^2-1\right) \omega  \left(M^2 \left(\left(3-2 k^2\right) u^2+3\right)+u^2 \left(k^4 u^2+k^2 \left(u^2+5\right)-3\right)-3 M^4\right)-u^3 \left(u^2-1\right) \omega ^2 \left(3 k^2 u^2-M^2+1\right)-3 i u^4 \left(u^2-1\right) \omega ^3\nn
\end{align}
}


\bibliographystyle{hunsrt}
\bibliography{DBIbib}


\end{document}